# A dynamic network approach for the study of human phenotypes

César A. Hidalgo[1, †], Nicholas Blumm[2,4], Albert-László Barabási[2,3,6], Nicholas A. Christakis[5]

[1]Center for International Development and Harvard Kennedy School, Harvard University, Cambridge, MA, 02138
[2]Center for Network Science, Dept. of Physics, Biology and Computer Science, Northeastern University, Boston, MA, 02115
[3]Center for Complex Network Research and Dept. of Physics, University of Notre Dame, Notre Dame, IN, 46556
[4]College of Computer and Information Science, Northeastern University, Boston, MA, 02115
[5]Department of Health Care Policy and Dept. of Medicine, Harvard Medical School, Boston, MA, 02115
[6]Department of Medicine, Harvard Medical School, Boston, MA 02115
[†] To whom correspondence should be addressed: cesar_hidalgo@ksg.harvard.edu

## ABSTRACT

**Background:**

*The use of networks to integrate different genetic, proteomic, and metabolic datasets has been proposed as a viable path toward elucidating the origins of specific diseases.*

**Methodology/Principle Findings**

*Here we introduce a new phenotypic database summarizing correlations obtained from the disease history of more than 30 million patients in a Phenotypic Disease Network (PDN). We present evidence that the structure of the PDN is relevant to the understanding of illness progression by showing that (1) patients develop diseases close in the network to those they already have; (2) the progression of disease along the links of the network is different for patients of different genders and ethnicities; (3) patients diagnosed with diseases which are more highly connected in the PDN tend to die sooner than those affected by less connected diseases; (4) diseases that tend to be preceded by others in the PDN tend to be more connected than diseases that precede other illnesses and are associated with higher degrees of mortality.*

**Conclusions/Significance**

*Our findings show that disease progression can be represented and studied using network methods, offering the potential to enhance our understanding of the origin and evolution of human diseases. The dataset introduced here, released concurrently with this publication, represents the largest relational phenotypic resource publicly available to the research community.*

# Author Summary

To help the understanding of physiological failures, diseases are defined as specific sets of phenotypes affecting one or several physiological systems. Yet, the complexity of biological systems implies that our working definitions of diseases are careful discretizations of a complex phenotypic space. To reconcile the discrete nature of diseases with the complexity of biological organisms, we need to understand how diseases are connected, as connections between these different discrete categories can be informative about the mechanisms causing physiological failures.

Here we introduce the Phenotypic Disease Network (PDN) as a map summarizing phenotypic connections between diseases and show that diseases progress preferentially along the links of this map. Furthermore, we show that this progression is different for patients with different genders and racial backgrounds and that patients affected by diseases that are connected to many other diseases in the PDN tend to die sooner than those affected by less connected diseases. Additionally, we have created a queryable online database ([http://hudine.neu.edu](http://hudine.neu.edu)) of the 18 different datasets generated from the more than 31 million patients in this study. The disease associations can be explored online or downloaded in bulk.

# INTRODUCTION

There are no clear boundaries between many diseases, as diseases can have multiple causes and can be related through several dimensions. From a genetic perspective, a pair of diseases can be related because they have both been associated with the same gene [1,2], whereas from a proteomic perspective diseases can be related because disease associated proteins act on the same pathway [3-9].

During the past half-decade, several resources have been constructed to help understand the entangled origins of many diseases. Many of these resources have been presented as networks in which interactions between disease-associated genes, proteins, and expression patterns have been summarized. For example, Goh et al. created a network of Mendelian gene-disease associations by connecting diseases that have been associated with the same genes [1] (see also Feldman et al. [2]), whereas Lee et al. constructed a network in which two diseases are linked if mutated enzymes associated with them catalyze adjacent metabolic reactions [10]. Network studies in the proteomic front have studied large protein interaction networks, like the ones created by Rual et al. [3] and Stelzl et al. [4], in an attempt to understand diseases like inherited ataxias [5] or Huntington's disease [6]. Moreover, in the gene expression front, microarray expression profiles and other cellular level information have been used to explore networks in inflammation [7], breast cancer [8], and brain disease [9].

While progress on the genetic and proteomic fronts has been impressive [1,10], much of the available resources overlook the fact that we have extensive and continually updated phenotypic information for humans – namely, patient clinical histories. Indeed, hospitals and insurance programs constantly collect detailed records for millions of patients. These datasets contain information on disease associations and progression. For example, such population-based disease associations could be used in conjunction with molecular and genetic data to help us uncover the molecular origins of diseases. Despite the potential utility of population based disease associations, extensive datasets linking diseases based on comorbidity associations do not exist, partly because access to extensive medical records is limited.

Typically, we say that a comorbidity relationship exists between two diseases whenever they affect the same individual substantially more than chance alone. One of our primary goals here is to make available pairwise comorbidity correlations for more than 10 thousand diseases reconstructed from over 30 million medical records. For completeness and utility, we organize the results in 18 different datasets. Each summarizes phenotypic associations extracted from four years worth of ICD9-CM claims data at the 5 and 3 digit level. Results are grouped into subsets of race, gender, and both race and gender (see SM). To facilitate their use, the datasets are available as a bulk download (http://hudine.neu.edu/resource/data/data.html ) or through a searchable web interface (http://hudine.neu.edu) that allows researchers, doctors and patients to explore these disease networks graphically, through an interactive Flash application, and numerically, by allowing them to generate tables summarizing the associations between a particular disease and all other diseases.

In the past, comorbidities have been used extensively to construct synthetic scales for mortality prediction [11,12], yet their utility could exceed their current use. Studying the structure defined by entire sets of comorbidities might help the understanding of many biological and medical questions from a perspective that is complementary to other approaches. For example, a recent study built a comorbidity network in an attempt to elucidate neurological diseases common genetic origins [13]. Heretofore, however, neither this data nor the data necessary to explore relationships between all diseases is currently available to the research community. Hence, here we decide to provide this data in the form of a Phenotypic Disease Network (PDN) capturing all diseases as recorded through medical claims. Additionally, we illustrate how a PDN can be used to study illness progression from a dynamic network perspective by interpreting the PDN as the landscape where illness progression occurs and show how the network can be used to study phenotypic differences between patients with different demographic backgrounds. Furthermore, we show that the local structure of a disease in the network, as characterized by its degree or number of connections, is associated with disease mortality. Finally, we study the directionality of disease progression, as observed in our dataset, and find that more central diseases in the PDN are more likely to occur after other diseases and that more peripheral diseases tend to precede other illnesses. We also find that patients diagnosed with diseases that tend to be preceded by other conditions tend to die sooner than those diagnosed with conditions that tend to precede other diseases. Together, these results and resources open new opportunities for biomolecular, bioinformatic and public health approaches to disease.

## METHODS

### Source Data and Study Population:

Hospital claims offer reliable, systematic, and complete data for disease detection [14,15,16]. Each record consists of the date of visit, a primary diagnosis and up to 9 secondary diagnoses, all specified by ICD9 codes of up to 5 digits. The first three digits specify the main disease category while the last two provide additional information about the disease. In total, the ICD-9-CM classification consists of 657 different categories at the 3 digit level and 16,459 categories at 5 digits. For a detailed list of currently used ICD9 codes see www.icd9data.com. We compiled raw Medicare claims [17,18] based on so-called MedPAR records regarding hospitalizations for 1990-1993. Medicare is the US government's health insurer, and it has information on 96% of all elderly Americans whether they seek medical care or not [19].

For the 32 million elderly Americans aged 65 or older enrolled in Medicare and alive for the entire study period, there were a total of 32,341,347 inpatient claims, pertaining to 13,039,018 individuals (the remaining individuals were not hospitalized at any point during this period). Demographically, our data set consists of patients over 65 years old (see Fig 1A for the age distribution) and is composed mainly of white patients, with a higher percentage of females (58.3% Fig 1B). Yet, the data set is large enough to estimate race and gender specific comorbity patterns.

## Data Limitations

The medical claims were made available to us is in the ICD-9-CM format, representing a controlled nomenclature constructed mainly for insurance claim purposes. Therefore, in some cases, more than one code corresponds to a particular disease, whereas in other cases codes are not specific enough for research purposes. For example, at the 5-digit level there are 33 diagnoses associated with hypertension, which reduce to five at the 3-digit level. Other times, the code is for a symptom such as "dehydration" which cannot be assigned to any one diagnosis. The vast majority of diseases, however, do map reliably to ICD9 codes.

While hospital claims have been proposed as a reliable method for disease detection [15,16,20], our data does not capture a complete cross section of the population. Our dataset consists of medical claims associated with hospitalizations of elderly citizens in the United States; thus, it contains limited information about diseases that are not common among elders from an industrialized country, such as many infectious diseases or pregnancy-related conditions. Nor does it contain information on patients who were not hospitalized and who instead seek solely outpatient care. Hence, it is important to interpret our results in the context of a population of elderly citizens in an industrialized country.

## Quantifying the Strength of Comorbidity Relationships

To measure relatedness starting from disease co-occurrence, we need to quantify the strength of comorbidities by introducing a notion of "distance" between two diseases (see Text S1). A difficulty of this approach is that different statistical distance measures have biases that over- or under-estimate the relationships between rare or prevalent diseases. These biases are important given that the number of times a particular disease is diagnosed –its prevalence- follows a heavy tailed distribution (Fig 1 C), meaning that while most diseases are rarely diagnosed, a few diseases have been diagnosed in a large fraction of the population. Hence, quantifying comorbidity often requires us to compare diseases affecting a few dozen patients with diseases affecting millions.

We will use two comorbidity measures to quantify the distance between two diseases: The Relative Risk ($RR$) and $\phi$-correlation ($\phi$). The $RR$ of observing a pair of diseases $i$ and $j$ affecting the same patient is given by

$$RR_{ij} = \frac{C_{ij} N}{P_i P_j} \quad (1)$$

where $C_{ij}$ is the number of patients affected by both diseases, $N$ is the total number of patients in the population and $P_i$ and $P_j$ are the prevalences of diseases $i$ and $j$. The distribution of $RR$ values found in our data set is shown in Fig 1 D. The $\phi$-correlation, which is Pearson's correlation for binary variables, can be expressed mathematically as:

$$\phi_{ij} = \frac{C_{ij} N - P_i P_j}{\sqrt{P_i P_j (N - P_i)(N - P_j)}} \quad (2)$$

The distribution of $\phi$ values representing all disease pairs where $C_{ij}>0$ is presented in Fig 1 E. A discussion on the confidence interval and statistical significance of these measures can be found in the Text S1.

These two comorbidity measures are not completely independent of each other (Fig 1 F), as they both increase with the number of patients affected by both diseases, yet both measures have their intrinsic biases. For example, *RR* overestimates relationships involving rare diseases and underestimates the comorbidity between highly prevalent illnesses, whereas $\phi$ accurately discriminates comorbidities between pairs of diseases of similar prevalence but underestimates the comorbidity between rare and common diseases (see SM Box 1). Given the complementary biases of the two measures, we construct a PDN separately for each measure and discuss their respective relevance to specific disease groups.

One important question is how the predictive power of comorbidity based relationships compares with that of heredity and known genetic markers. Of the two measures discussed above, the Relative Risk *(RR)* enjoys the most widespread use in the medical literature [21-30], making it the most suitable for such comparison. We find that the relative risk of being diagnosed with one disease given another disease affecting a patient in our data varies in the range *RR*~0.25-16 (Fig 1 D). Sibling studies have found that the relative risk of having a disease given that a sibling has the same disease typically ranges from *RR~3* for type 2 diabetes [21] to *RR~2-7* for early myocardial infarction [22], *~7-10* for bipolar disorder [23,24] and rheumatoid arthritis [25] and *~17-35* for Crohn's Disease [26]. Most of these values fall in the range of relative risks associated with our observed comorbidities. Hence, statistically speaking, the magnitude of the disease risk predicted by comorbidity relationships is comparable to that of family history. Furthermore, we can compare comorbidity statistics with typical relative risk values found in genetic susceptibility studies. For example, the relative risk of type 2 diabetes for carriers of the at-risk allele TCF7L2 ranges between *RR~1.45* and *2.41* [27], whereas the rs2476601 SNP in the PTPN22 gene confers a genetic relative risk for rheumatoid arthritis of *RR~1.8* [28,29]. In contrast, the *RR* for a type 2 diabetes of a patient diagnosed with Ischemic Heart Disease is *RR~1.61*, whereas a rheumatoid arthritis patient is at *RR~3.64* for the disease if he or she is diagnosed with osteoporosis [30]. The statistical strength of the observed comorbidities is therefore comparable to that found in siblings and genetic susceptibility studies, a favorable comparison that provides further motivation to use comorbidity data to explore disease risk.

# RESULTS

## The Phenotypic Disease Network

We can summarize the set of all comorbidity associations between all diseases expressed in the study population by constructing a Phenotypic Disease Network (PDN). In the PDN, nodes are disease phenotypes identified by unique ICD9 codes, and links connect phenotypes that show significant comorbidity according to the measures introduced above.

In principle, the number of disease-disease associations in the PDN is proportional to the square of the number of phenotypes, yet many of these associations are either not strong or are not statistically significant (see SM). Hence, we explore the structure of the PDN by focusing on the strongest and most significant of these associations. To achieve this, we offer two visualizations of the PDN (see SM), the first constructed using $RR$ (Fig 2 A) and the second using $\phi$ (Fig 2B and Text S1).

While there are many similarities between the two networks, such as the proximity between nephritis and hypertension or psychiatric disorders and poisoning, the overall structure of the PDN and the specific disease groups present in each one of them reflect the individual biases of the metric used to construct the links. The network constructed using $RR$ (Fig 2 A) is populated by relatively infrequent illnesses and has visually discernable modules that follow the ICD9 classification somewhat closely. In contrast, the network constructed using $\phi$ (Fig 2 B) is populated by highly prevalent diseases with many connections across different ICD9 categories. Despite these differences between the two networks, we do not argue in favor of one particular representation; they both capture statistically significant associations at different prevalence scales. Together, each offers a complementary representation of the phenotypic disease network.

## Disease Network Dynamics

While a network representation of diseases has many potential applications, here we concentrate on three examples illustrating the use of the PDN to study the illness progression from a network dynamics perspective [31]. The PDN can be seen as a "map" of the phenotypic space. This map allows us to study illness progression as a dynamic network process in which patients "jump" from one disease to another along the links of the PDN [31]. Our ability to fully develop such a view of diseases is limited, however, by our data. While we can order diseases according to the date they were diagnosed, we cannot exclude the possibility that the observed progression is a result of our limited observation window. For example, a patient in our data set can be diagnosed with type II diabetes on the first visit and hypertension on the second visit. Yet, lacking information on previous disease history, we cannot conclude that diabetes precedes hypertension, as hypertension could have been diagnosed at any earlier time point not recorded in our data. Hence we begin our analysis using a conservative approach in which we study possible consequences of disease progression in a static network picture and continue, by the end of the paper, to study the observed directionality of disease progression, limiting any conclusions due to the aforementioned biases.

These limitations require us to adopt a more conservative approach in our analysis. Here we explore disease network dynamics by asking three questions (Fig 3 A). $Q_1$: Does illness significantly progress along the links of the PDN? $Q_2$: Is illness progression different for patients of different races and genders? $Q_3$: Does the connectivity of a disease, as measured in the PDN, correlates with higher lethality?

To answer the first question ($Q_1$) we use a recently introduced method to decide whether a node property spreads along the links of a network [31] (see Text S1). We measure the average correlation between diseases diagnosed in the first two visits and those diagnosed in the 3$^{rd}$ and 4$^{th}$ visits for all patients with four visits (*N=946,580*). We controlled for the correlations inherent to the dataset by repeating the procedure using a randomized set of diseases for the first two visits extracted in such a way that the prevalence of each disease in the randomized sets matches the one observed in the original data. We find that diseases diagnosed in the first two visits are more correlated with those diagnosed in the last two visits than what we observe on our control case (Fig 3 B & C). In a case-by-case basis, we can compare the correlations between the real and randomized measures by calculating the ratio $H = \bar{\phi}_D / \bar{\phi}_C$, where $\bar{\phi}_D$ is the average correlation between the diagnoses received by a patient in his first two and last two visits and $\bar{\phi}_C$ is the average correlation found in the control case. The distribution of *H* (Fig 3 D & E) reveals that inter-visit correlations are larger than would be expected by chance for 95.6% of the patients on the $\phi$-PDN and for 81.5% of the patients in the *RR*-PDN (by a factor of 10 on average for the $\phi$-PDN and of 1.5 for the *RR*-PDN). Therefore, it is a valid approximation to think of the development of patients' illnesses as a spreading process over a PDN. We note that while the effect discussed above is present for both the $\phi$ and *RR*-based networks, it is more pronounced in the $\phi$-PDN, suggesting a superior potential predictive power for the $\phi$-representation. We find that this result is not affected by including the individuals with four visits or not in the PDN (see Text S1). This is because of the large number of observations and the regularity of the observed disease correlations.

While our data does not allow us to be conclusive about the directionality of disease progression, differences in the strength of comorbidity relationships can still indicate differences in the dynamics of illness progression. The reason is that patients affected by a pair of diseases had traversed the link between them at some point in time and in one of the two possible directions. Here, we explore $Q_2$ by looking at differences in the strength of the observed comorbidities for patients from different ethnic background and genders. For this we calculate the odds ratio for the difference in comorbidity between diseases *i* and *j* as expressed in populations $\alpha$ and $\beta$. Mathematically, the odds ratio is defined as

$$OR_{ij}(\alpha,\beta) = \frac{p_{ij}(\alpha)(1-p_{ij}(\beta))}{p_{ij}(\beta)(1-p_{ij}(\alpha))} \ , \qquad (3)$$

where $p_{ij}(\alpha) = C_{ij}(\alpha)/N_\alpha$ is the probability that diseases *i* and *j* are observed in a patient of population $\alpha$.

We discuss as an example a network showing differences in the strength of comorbidities between white and black males. We illustrate this on the subset of Figure 2 B in which all diseases connected to Hypertension or Ischemic Heart Disease are shown. In Figure 3 F, blue links connect diseases that are significantly more comorbid for black males, whereas red links connect diseases that are more comorbid for white males. This picture suggest that ischemic heart disease, infarctions, hypercholesterolemia, and pulmonary complications, among other diseases, tend to be more comorbid in white males than in black males; whereas hypertension, diabetes, and renal and other disorders tend to be more comorbid in black males than in white males. The structure presented in Figure 3 F summarizes well known disease associations, validating the ability of the PDN to explore gender and race variations on comorbidity, which could help discern disease etiology. Figure 5S shows a similar example for males and females. Comparative studies like this one can be performed for any disease using the project's website (http://hudine.neu.edu).

Finally, we explore our third question ($Q_3$) by showing that the lethality of a disease is associated with its connectivity in the PDN. We can quantify the connectivity of a particular disease by adding the correlations between a disease and all other diseases to which it is connected [32,33]. We use $K_i^\phi = \sum_j \phi_{ij}$ and $K_i^{RR} = \sum_j RR_{ij}$ respectively for the $\phi$- and RR-networks. Both $K_i^\phi$ and $K_i^{RR}$ tell us how embedded disease *i* is in the PDN; high values of $K_i^\phi$ and $K_i^{RR}$ indicate that disease *i* is strongly connected to many other diseases in the PDN. To measure the lethality of a disease, we calculated the percentage of people deceased within the 8 years following the first diagnosis recorded in our database. We find that disease connectivity and lethality are correlated for the $\phi$-PDN and the *RR*-PDN (Fig 4 A & B). A simpler and contrasting hypothesis is to test whether the lethality of a disease correlates with its prevalence (Fig 4 C); we find that prevalence shows only a weak correlation with lethality and cannot explain the effect seen in Fig 4 A & B. We also find that the strength of the relationship between disease connectivity and lethality is greater for some groups of diseases than others (Table S3). For example, this relationship is strong for neoplasms (Fig 4 D & E) whereas for mental disorders the correlation is week (Fig 4 F) or even negative (Fig 4 G).

A possible explanation for the observed correlation between connectivity and lethality is that sicker patients accrue more diagnoses and hence the observed correlation is just a restatement of this trivial fact. We can rule this out by looking at the correlation between the average connectivity of diseases diagnosed to patients with a given number of hospital visits, diagnoses, and number of years they remained alive after the last diagnosis was observed. We performed this analysis by looking at data on the 7,878,255 patients for which we know the exact year of death; the remaining patients were reported as either alive or unknown in our data set. Figures 5 A and 5 B show the histogram for the number of visits and diagnoses assigned to this set of 7,878,255 patients. Figures 5 C and 5 D show that there is a significant negative correlation between the average connectivity of diseases observed in patients with the same number of visits or diagnoses or the number of years they survived after the last diagnosis was observed. Hence the observed correlation between connectivity and lethality does not come from a simple accumulation of diagnoses by sicker patients. Our results indicate that the severity of a disease can be approximated by its connectivity in the PDN for patients with the same number of diagnoses, hence the structure of the PDN matters as the location of a patient in the PDN is a predictor of the number of years he is expected to remain alive. We also notice that in this example the PDN created using $\phi$ correlates more strongly with the number of years a person survived than the RR-PDN. In Text S1, we show that the observed correlation between the connectivity in the PDN of diseases affecting a patient and the number of years survived are robust to simultaneous control for age, gender, number of diagnoses, and number of visits.

Finally, we briefly analyze the directionality of disease progression, as observed in our data, keeping in mind that the limited observation period of our study limits our ability to be conclusive about disease directionality because of the aforementioned reasons. Hence, we interpret the following results as suggestive evidence of directionality rather than as a proof. To reduce the noise levels of our analysis we concentrate on links between diseases affecting at least 1 in 500 patients (0.2%), which from the size of our data set, are expected to co-occur in at least 50 patients. At the 5 digit level our comorbidity data contains 133,858 links connecting the 518 diseases affecting at least 1 out of 500 patients.

Consider the link connecting diseases *i* and *j*. To assign a direction to this link we begin counting the number of times disease *i* was diagnosed before disease *j* and represent this number as $L_{i \to j}$. When computing $L_{i \to j}$ we disregard those cases in which both diseases were diagnosed for the first time in the same visit, as our data does not allow us to study precedence within the same hospitalization; hence $L_{i \to j} + L_{j \to i} \leq C_{ij}$. Most links connect diseases with large differences in prevalence; hence we normalize $L_{i \to j}$ by the prevalence of the disease *i* using the formula $l_{i \to j} = (L_{i \to j} + 1)/P_i$, where the factor of one is added to include, when taking ratios, those cases in which $L_{i->j}$ is equal to zero. In such cases $l_{i->j}=1/P_i$. We introduce this normalization because the probability that a disease is diagnosed before another disease is proportional to its prevalence. Finally, we can assign a direction to a link by creating a variable that, after controlling for the prevalence of a disease, is positive if a disease tends to precede another disease (outgoing link) and negative if a disease tends to come after the disease at the other end of that link (incoming link). We define the *directionality* $\lambda_{i->j}$ of the link connecting disease *i* to disease *j* as:

$$\lambda_{i->j} = \log_{10}\left(\frac{l_{i->j}}{l_{j->i}}\right) \quad (4)$$

A value of $\lambda_{i->j}=1$ indicates that, after controlling for prevalence, the probability a patient is diagnosed with disease *i* before it is diagnosed with disease *j* is 10 times higher than the probability a patient is diagnosed with disease *j* before being diagnosed with disease *i*. Whereas a value of $\lambda_{i->j}=2$ indicates that the ratio between this probabilities is equal to 100. Fig 6 A shows the distribution of $\lambda_{i->j}$ calculated for the 133,858 links connecting diseases affecting at least 1 out of 500 patients. We find that this distribution has a well defined peak close to $\lambda_{i->j}=0$, indicating that the most common type of link is that without a preferred direction. Despite this, there are a substantial number of links that do appear to show a preferred direction. We find that, out of the 133,858 links considered, 15,625 (11.7%) of them lie outside the $\lambda_{i->j}$ ]-1,1[ whereas only 229 (0.2%) lie outside the $\lambda_{i->j}$ ]-2,2[ interval.

The directionality analysis allows us to extend our study of disease connectivity and lethality to include the directionality of the links connecting a disease to other diseases in the PDN. By assigning a direction to the links connecting a disease with other diseases in the PDN we can classify diseases into *source* and *sink* diseases; *source* diseases being those whose links are more likely to point away from them and *sink* diseases being those whose links are more likely to point towards them. To capture this effect we define the *precedence* of disease *i* as the sum of the directionality of the links connecting a disease to all of its neighboring diseases in the PDN:

$$\Lambda_i = \sum_j \lambda_{i->j} \qquad (5)$$

$\Lambda_i$ is positive for diseases that tend to come before other diseases and is negative for diseases that tend to come after other diseases. We find that $\Lambda_i$ is not independent of disease prevalence, as it exhibits a slow, logarithmic, dependence on it (Figure 6 B). We can remove the dependence of $\Lambda_i$ in the prevalence of a disease by subtracting the trend directly from it. This allows us to obtain a detrended measure of disease precedence ($\Lambda_i^*$, Figure 6 B inset) which is independent of prevalence and can be used to explore the information on lethality contained in the structure of the PDN.

Figure 6 C shows that $\Lambda_i^*$ is negatively correlated with the connectivity of a disease, indicating that highly connected diseases in the PDN tend to come after other diseases, rather than before, suggesting that highly connected diseases more likely represent advanced stages of illness. Finally, we study the relationship between disease precedence and lethality, finding that patients diagnosed with sink diseases tend to die sooner than those diagnosed with source diseases, as measured from our directionality analysis in the PDN. We checked whether this result was just a restatement of our previous finding, indicating that patients diagnosed with highly connected diseases tend to die sooner than those diagnosed with sparsely connected diseases, and found that both effects are simultaneously significant (see SM). Furthermore, our statistical analysis shows that for relatively short terms (2 years) disease precedence is a better predictor of lethality than disease connectivity, whereas disease connectivity appears to be a better predictor of lethality than disease precedence for longer terms (8 years). Hence both, the connectivity and precedence of a disease, carry important information on the burden that a given disease signifies for patients affected by it.

# DISCUSSION

While there is a great deal of expectation that disease associations are of enormous potential value to the research community, the lack of phenotypic data available to complement genotypic and proteomic datasets has limited scientific progress towards elucidating the origins of human disease. Here we take a step toward rectifying this situation by introducing an extensive, publicly available data set quantifying comorbidity associations expressed in a large population.

An important issue raised by calls for phenotypic network information is the potential integration of phenotypic data with genetic and proteomic data to better elucidate disease etiology. There are, however, other potential applications of a network-based approach to diseases. Phenotypic "maps" like the ones presented here could be used to study the disease evolution of patients and represent an ideal way to visualize and represent medical health records in a future in which digital medical records will need to be accessed by health care workers in a delocalized manner [34].

Here we have shown suggestive evidence that patients develop diseases close in the PDN to those already affecting them. We also showed that the PDN has a heterogeneous structure where some diseases are highly connected while others are barely connected at all. While not conclusive, these observations can explain the observation that more connected diseases are seen to be more lethal, as patients developing highly connected diseases are more likely those at an advanced stage of disease, which can be reached through multiple paths in the PDN.

Exploring comorbidities from a network perspective could help determine whether differences in the comorbidity patterns expressed in different populations indicate differences in biological processes, environmental factors, or health care quality provided for each population. Here we show as a first step that there are differences in the strength of co-morbidities measured for patients of different races and gender. The PDN could be the starting point of studies exploring these and related questions. This is why we make our data available to the research community at (http://hudine.neu.edu)

**Acknowledgements**

We thank Laurie Meneades for the expert data programming required to build the analytic data set and Z. Oltvai and C. Teutsch for useful medical discussions.

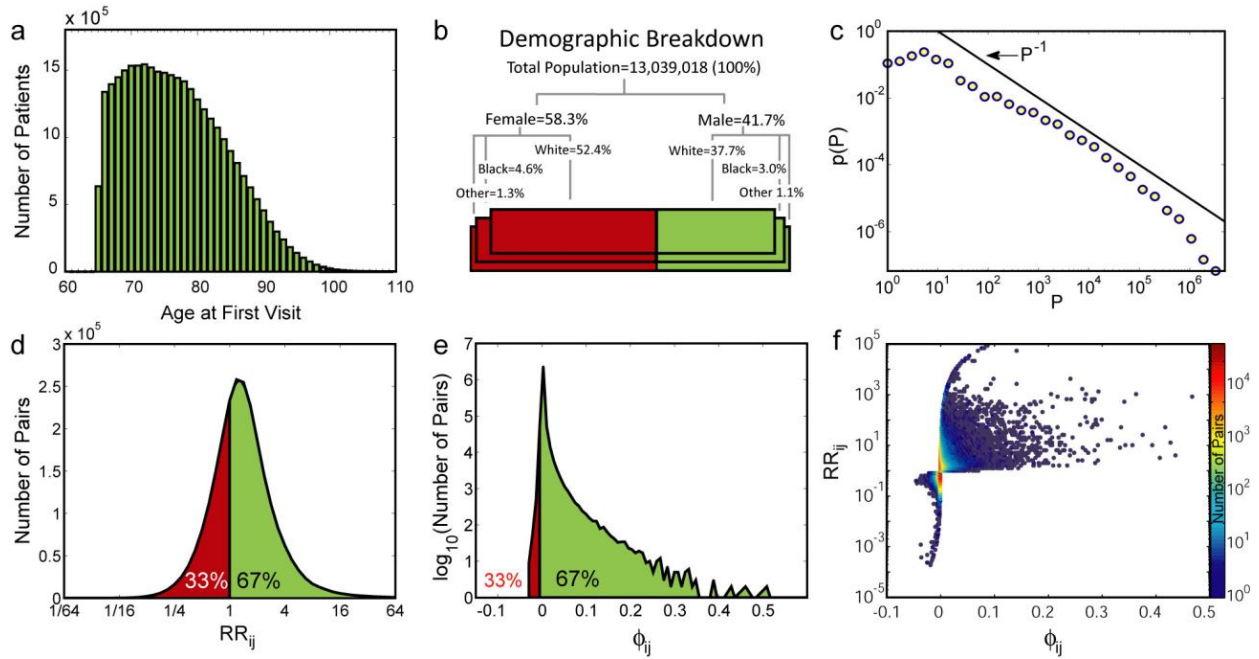

**Figure 1** Data characteristics and basic comorbidity statistics. **A** Age distribution for the study population. **B** Demographic breakdown of the study population. **C** Prevalence distribution for all diseases measured using ICD9 codes at the 5 digit level. **D** Distribution of the relative risk (*RR*) between all disease pairs. **E** Distribution of the $\phi$-correlation between all disease pairs. **F** Scatter plot between the $\phi$-correlation and the relative risk of disease pairs.

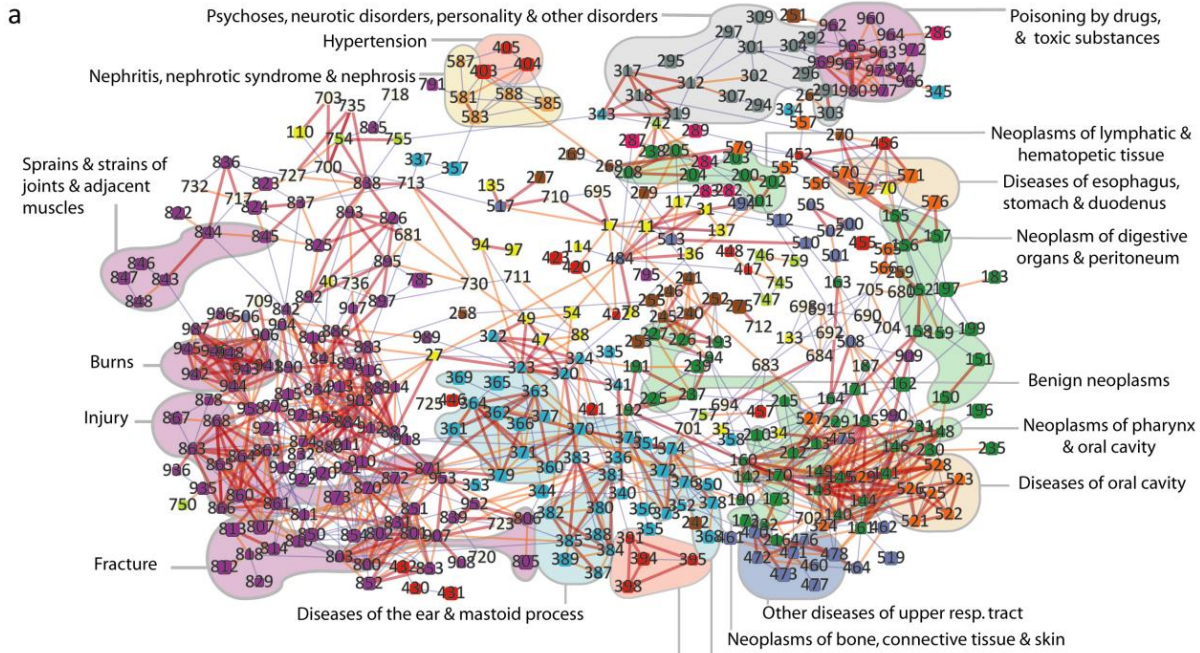
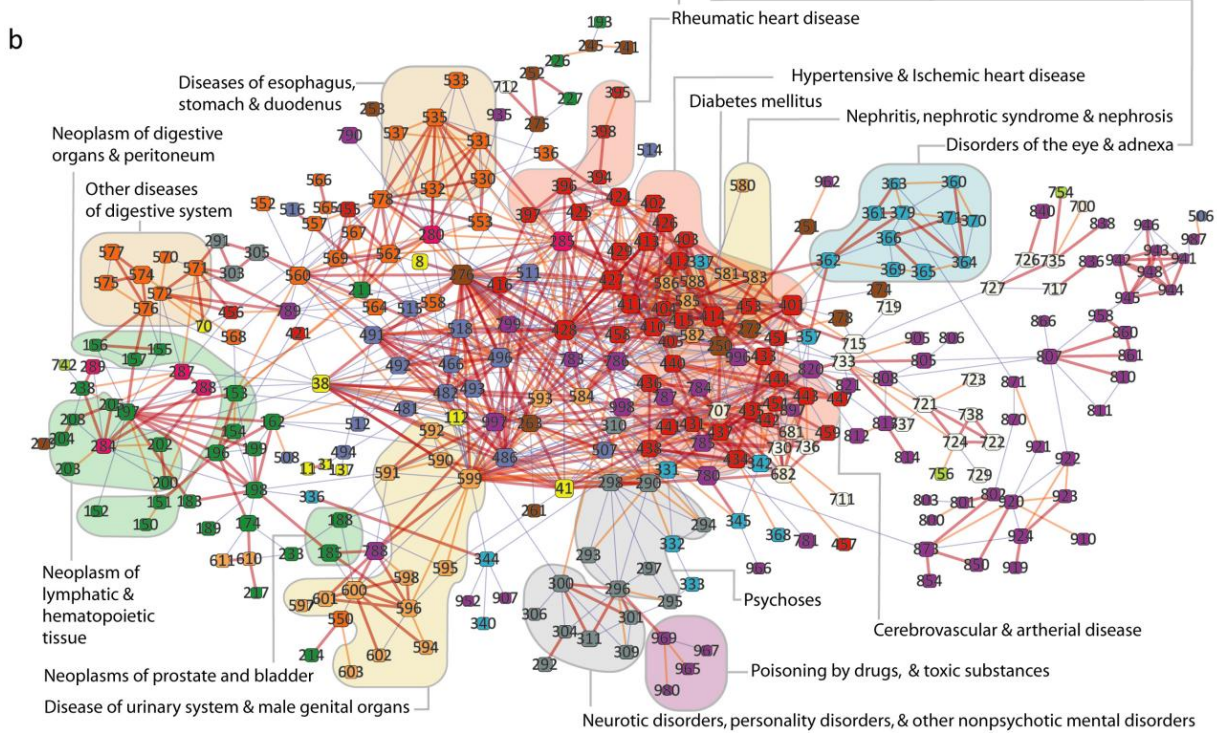

**Figure 2**. Phenotypic Disease Networks (PDNs). Nodes are diseases; links are correlations. Node color identifies the ICD9 category; node size is proportional to disease prevalence. Link color indicates correlation strength. **A** PDN constructed using *RR*. Only statistically significant links with $RR_{ij} > 20$ are shown. **B** PDN built using $\phi$-correlation. Here all statistically significant links where $\phi > 0.06$ are shown.

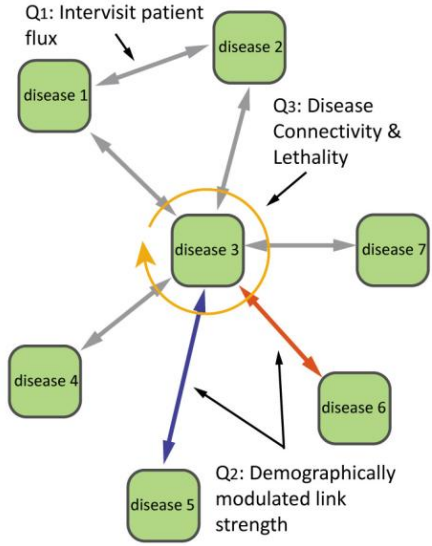
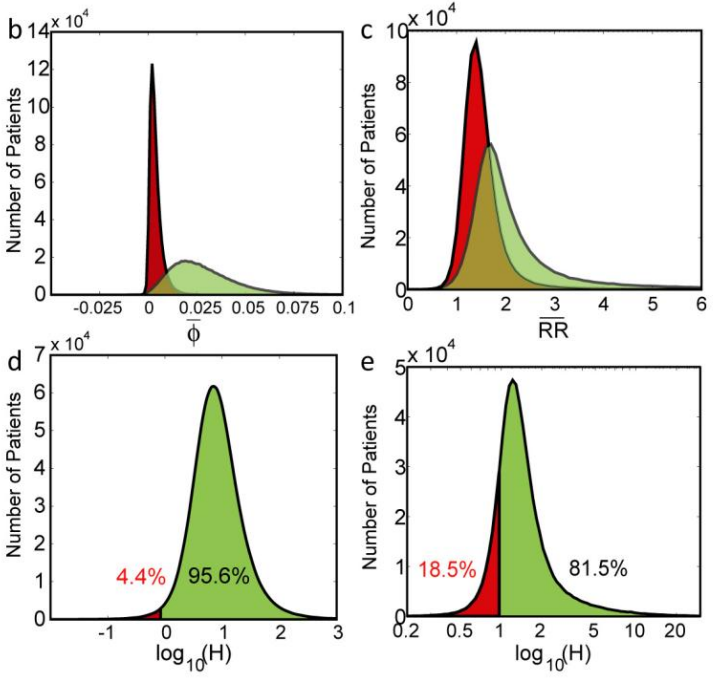
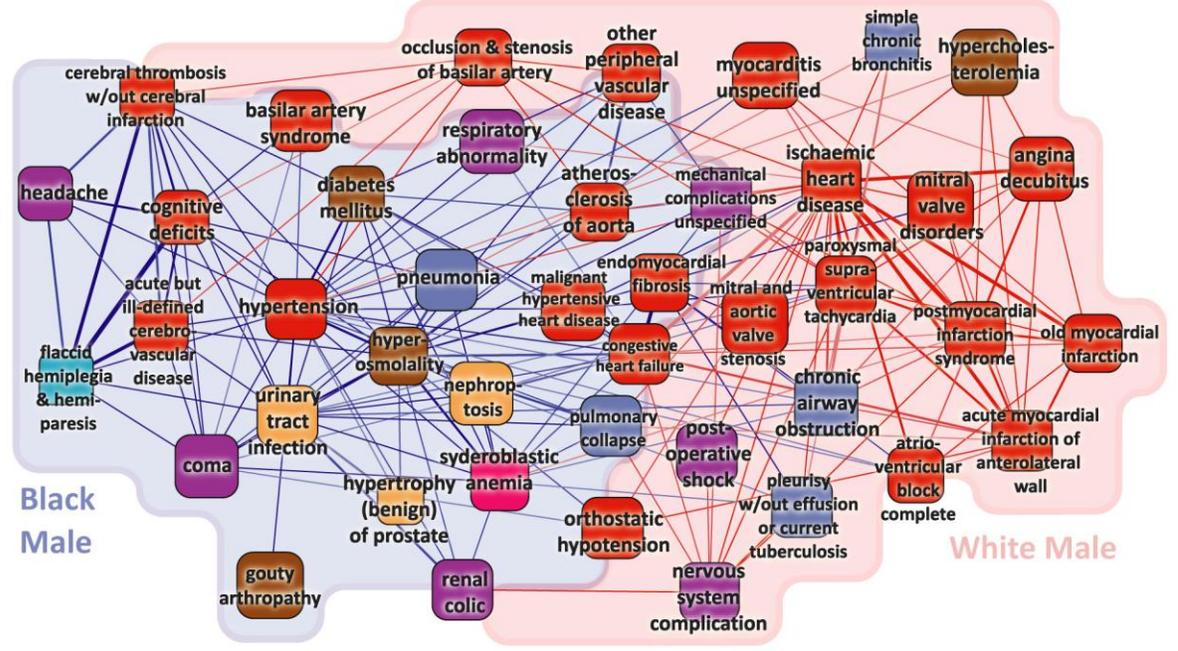

**Figure 3** The Phenotypic Disease Network and disease dynamics. **A** Schematic representation of the three dynamical questions explore here. **B** Average $\phi$-correlation between diseases diagnosed in the first two and last two visits for the 946,580 patients with 4 visits (green) and when we consider a randomized set of diseases for the first two visits (red). **C** same as **B** but for the *RR*-PDN. **D**. Ratio between the average $\phi$-correlation among diagnoses received by a patient in its first two and last two visits relative to the control case. **E**. same as **D** but for the *RR*-PDN. **F**. Gender and race differences. The subset of Fig 2 **B** where all diseases connected to hypertension and ischemic heart disease is shown. Blue links indicate comorbidities that are strongest among black males; whereas red links indicate comorbidities that are strongest among white males (see legend).

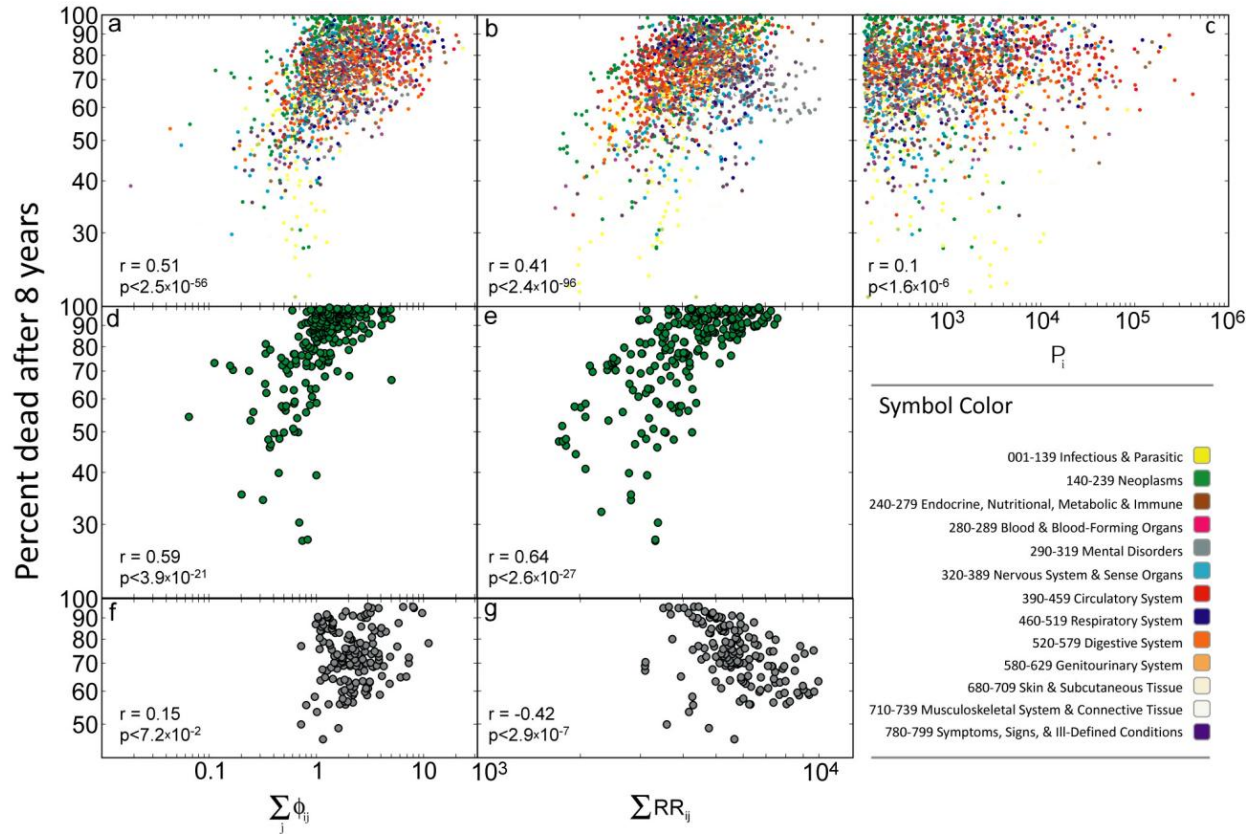

**Figure 4** Disease connectivity and lethality. **A** Scatter plot between the connectivity of a disease measured in the $\phi$-PDN and the percent of patients that died 8 years after this disease was first observed in our data set. **B** Same as **A** for the *RR*-PDN **C** percent of patients that died 8 years after this disease was first observed in our data set as a function of disease prevalence. **D** same as **A** showing only neoplasms. **E** same as **B** showing only neoplasms. **F** same as **A** showing only mental disorders. **G** same as **B** showing only mental disorders.

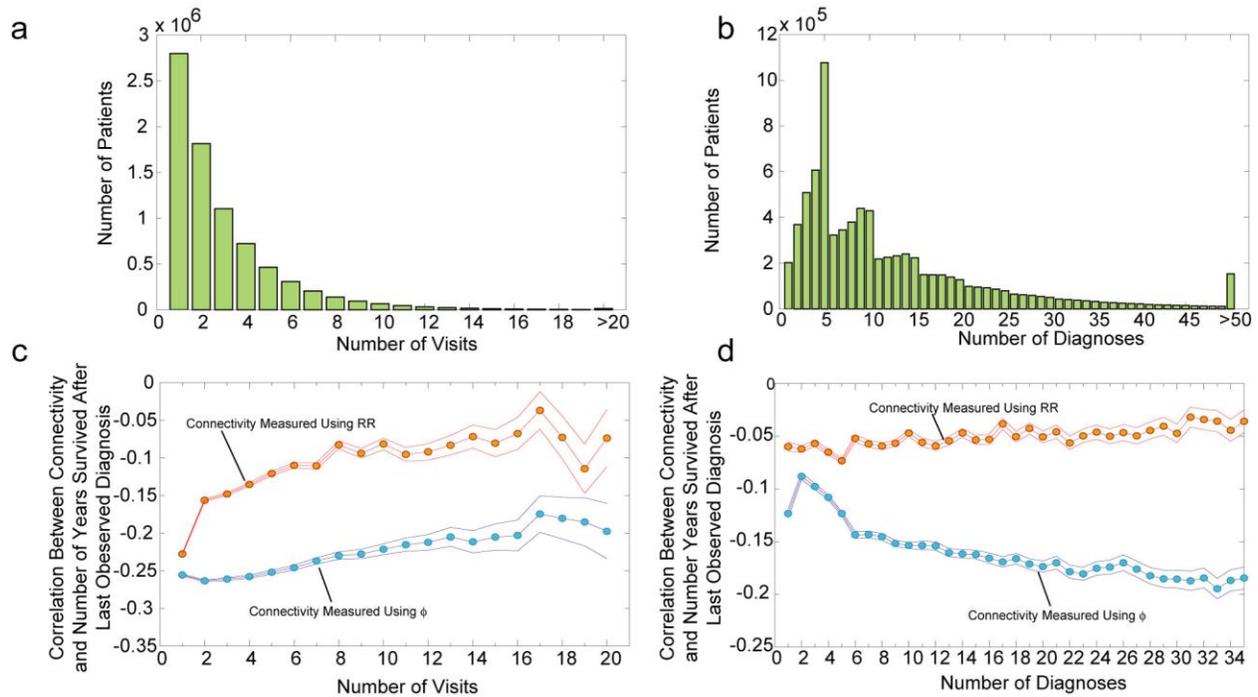

**Figure 5** Connectivity lethality control. **A**. Histogram with the number of visits for each patient for which the year of death is known. **B.** Histogram for the number of diagnosis assigned to each patient for which the year of death is known. **C.** Correlation between the average connectivity of the diagnosis assigned to a patient and the number of years survived after the last diagnosis was recorded for groups of patients with the same number of hospital visits. **D.** Correlation between the average connectivity of the diagnosis assigned to a patient and the number of years survived after the last diagnosis was recorded for groups of patients with the same number of total number of diagnosis assigned. Error margins in **C** and **D** represent 95% confidence intervals.

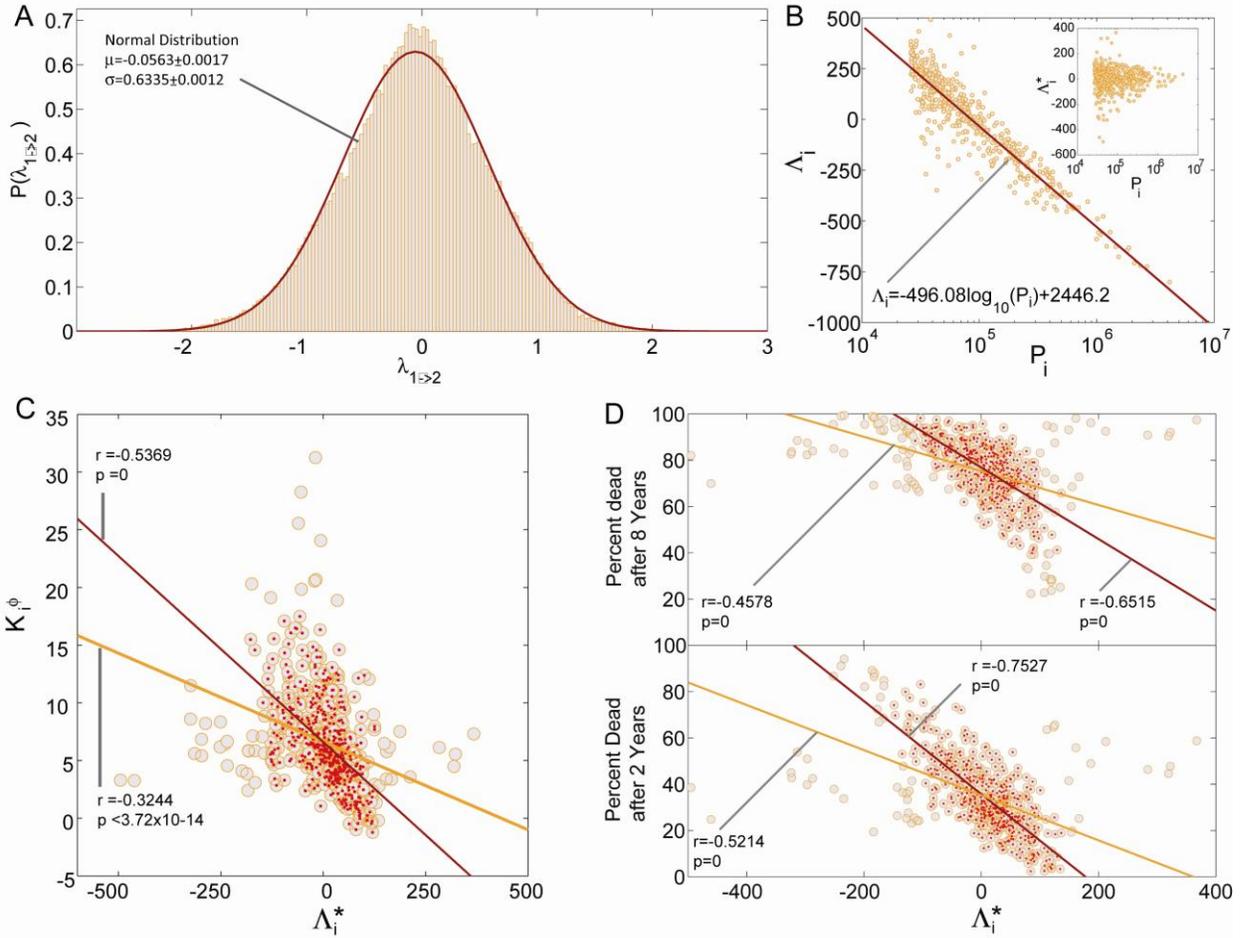

**Figure 6** Directionality of disease progression. **A.** Distribution of $\lambda_{1\to 2}$ **B.** Disease precedence $\Lambda_i$ as a function of disease prevalence $P_i$. The inset shows the same plot after removing the trend from disease precedence ($\Lambda_i^* = \Lambda_i + 496.08\log_{10}(P_i) - 2446.2$) **C.** Disease connectivity calculated from the ϕ-PDN as a function of $\Lambda_i^*$. The green line shows the best fit for the 518 diseases with a prevalence larger than 1/500 (green circles) while the red line shows the best fit for the 463 diseases at the center of the cloud (red points). The correlation coefficient is represented by r and its associated p-value by p. **D.** Percentage of patients that died 2 and 8 years after being diagnosed with a disease with a given detrended precedence $\Lambda_i^*$. The green lines show the best fit for all the 518 diseases (green circles) while the red lines show the fit for the 434 (top panel) and 465 (bottom panel) diseases at the bulk of the cloud.